\documentclass[twoside,12pt]{article}
\usepackage{epsfig}

\newcommand{\be}{\begin{equation}}
\newcommand{\ee}{\end{equation}}
\newcommand{\bea}{\begin{eqnarray}}
\newcommand{\eea}{\end{eqnarray}}

\begin{document}

\title{ \vspace{1cm} Anisotropies in momentum space at finite Shear Viscosity in ultrarelativistic
heavy-ion collisions}
\author{V. Greco, $^{1,2}$ M. Colonna, $^1$ M. Di Toro, $^{1,2}$ G. Ferini $^2$ 
\\
$^1$INFN-LNS, Laboratori Nazionali del Sud, Italy\\
$^2$ Dipartimento di Fisica e Astronomia, Universit\'a di Catania, Italy\\
}
\maketitle
\begin{abstract}
Within a parton cascade we investigate the dependence of anisotropies in momentum space, namely the elliptic flow
$v_2=<cos(2\phi)>$ and the $v_4=<cos(4\phi)>$, on both the finite shear viscosity $\eta$ and the freeze-out (f.o.) dynamics
at the RHIC energy of 200 AGeV. In particular it is discussed the impact of the f.o. dynamics looking at two different procedures: switching-off the collisions when the energy density goes below a fixed value or reducing the cross section according to the increase in $\eta/s$ from a QGP phase to a hadronic one.
We address the relation between the scaling of $v_2(p_T)$ with the eccentricity $\epsilon_x$ and with the integrated elliptic flow. We show that the breaking of the $v_2(p_T)/\epsilon_x$ scaling is not coming mainly from the finite $\eta/s$ but from the
f.o. dynamics and that the $v_2(p_T)$ is weakly dependent on the f.o. scheme. On the other hand the $v_4(p_T)$ 
is found to be much more dependent on both the $\eta/s$ and the f.o. dynamics and hence is indicated to put better
constraints on the properties of the QGP. A first semi-quantitative analysis show that both $v_2$ and $v_4$ (with
the smooth f.o.) consistently indicate a plasma with $4\pi \eta/s \sim 1-2$. 
\end{abstract}
\section{Introduction}

The first stage at the Relativistic Heavy Ion Collider (RHIC) has successfully shown the formation of hot and dense matter which behaves like a nearly perfect fluid. Such a conclusion mainly relies on the large value of the elliptic flow $v_2$ \cite{Kolb:2003dz,Ackermann:2000tr}. However there is a growing evidence of a breakdown of the ideal hydrodynamical behavior especially in the intermediate $p_T$ region ($1.5<p_T<5$ GeV) \cite{Abelev:2007rw,greco,Annual} and in peripheral collisions that calls for
a kinetic approach to study finite shear viscosity and non-equilibrium effects. 
A currently debated topic is in particular the effect of the conjectured minimal shear viscosity to entropy density ratio $\eta/s \geq 1/4\pi$ \cite{Kovtun:2004de} on collective flows. 
Several exploratory studies within viscous hydrodynamics\cite{Drescher:2007cd,Romatschke:2007mq,Song:2007fn,Song:2008si} or cascade approaches \cite{Xu:2007jv,Ferini:2008he,Molnar:2008jw} has shown a significant effect of the finite $\eta/s$ even for values around the lower bound, indicating a value for $\eta/s \sim 0.1-0.2$.

The analysis presented here is based on a parton cascade approach for massless
particles and it is mainly focused on the intermediate $p_T$ region.
The main idea is to keep the $\eta/s$
of the medium constant during the collision dynamics by rearranging the parton cross section according to the local density and mean momentum values as suggested by D. Molnar in \cite{Abreu:2007kv}.
We study the scaling of the $v_2(p_T)$ with the eccentricity 
$\epsilon_x =\langle y^2 -x^2\rangle/\langle x^2+y^2\rangle$ as a function of the centrality of the collision
for $Au+Au$ collisions at $\sqrt{s_{NN}}= 200$ GeV.
We show that if $\eta/s$, in the range $1<4\pi \eta/s<4$, is kept constant down to the thermal
freeze-out ($\epsilon \sim 0.2$ GeV/fm$^3$) the scaling is not broken by the finite shear
viscosity in the whole $p_T$ range investigated (up to 3.5 GeV). 
Hence the scaling is not a unique feature of ideal hydrodynamics \cite{Bhalerao:2005mm}.
On the other hand experimentally it has been observed a scaling of $v_2(p_T)/\langle v_2\rangle$ \cite{Adare:2006ti}
with the centrality of the collision and the system size together with the broken scaling of
$v_2(p_T)/\epsilon_x$ \cite{:2008ed}. 
The question is what is the origin of the breaking once a value of $\eta/s \sim 0.1-0.2$ is not found
to be responsible for it. In Ref. \cite{Ferini:2008he} we pointed out that once a freeze-out condition 
is introduced at $\epsilon_{f.o.} \sim$ 0.7 GeV/fm$^3$
a cascade approach at finite viscosity can account for the breaking of the
scaling for $v_2(p_T)/\epsilon_x$ together with a persisting scaling for $v_2(p_T)/\langle v_2\rangle$, as 
experimentally observed. 
Here we present a further developments of such a study replacing the cut-off freeze-out based on a smooth
increase of the $\eta/s$ in the region of the mixed quark-hadron phase ($\epsilon \sim 0.3-1.7$ GeV/fm$^3$).
We see that the $v_2(p_T)$ is similar in the two freeze-out prescriptions confirming our first results \cite{Ferini:2008he}.
Moreover we analyze for the first time the effect of finite $\eta/s$ on the $v_4$ that is revealed to
be much stronger than for the elliptic flow and much more sensitive to the way the system is 
described at energy densities $\epsilon \sim 0.5-2$ GeV/fm$^3$.
We finally emphasize that the shape of $v_2(p_T)$  and $v_4(p_T)$ at $4\pi\eta/s \sim 2$ is consistent with
to the one conjectured in coalescence
models \cite{Annual}, hence a definitive evaluation of $\eta/s$ is entangled with the observation of quark number scaling
in the same $p_T$ range. Hadronization by coalescence plus fragmentation has to be self-consistently
included in the next future.

\section{The parton cascade}
Our approach to study the effect of finite shear viscosity on the development of anisotropies in momentum space
is based on a $3+1$ dimensional Montecarlo cascade \cite{Ferini:2008he} for on-shell partons
based on the stochastic interpretation of the transition rate thoroughly discussed in Ref.\cite{Xu:2004mz}. 
Therefore the evolution of parton distribution function from initial conditions is 
governed by the Boltzmann equation 
\begin{equation}
 p_{\mu} \partial^\mu f_1 \!=\! 
\int\limits_2\!\!\! \int\limits_{1^\prime}\!\!\! \int\limits_{2^\prime}\!\!
 (f_{1^\prime} f_{2^\prime}  -f_1 f_2) \vert{\cal M}_{1^\prime 2^\prime \rightarrow 12} \vert^2 
 \delta^4 (p_1+p_2-p_1^\prime-p_2^\prime)
\end{equation}
where $\int_j= \int_j d^3p_j/\left[ (2\pi)^3\, 2E_j \right] $, $\cal M$ denotes the transition matrix for the elastic processes and $f_j$ are the particle distribution functions.
For the numerical implementation, we discretize the space into cells small respect to the system size and 
we use such cells to calculate all the local quantities. 
Several checks have been performed as in \cite{Xu:2004mz} to test the validity of the code and
to choose a good discretization for the convergence of the results for momentum anisotropies.

In kinetic theory under ultra-relativistic conditions the shear viscosity can be expressed as \cite{degroot}
\begin{equation}
 \eta=\frac{4}{15} \rho \langle p \rangle \lambda
\label{eq1}
\end{equation}
with $\rho$ the parton density, $\lambda$ the mean free path and $\langle p \rangle$ the average momentum.
Therefore considering that the entropy density for a massless gas is $s=\rho(4-\mu/T)$, $\mu$ being
the chemical potential, we get:
\begin{equation}
 \eta/s=\frac{4}{15} \frac{\langle p \rangle}{\sigma_{tr} \rho (4 - \mu/T )}
\label{eq2}
\end{equation}
where $\sigma_{tr}$ is the transport cross section, i.e. the $\texttt{sin}^2 \theta$ weighted cross section.  
We use a pQCD inspired cross section with the infrared singularity regularized by 
Debye thermal mass $m_D$ \cite{moln02}:
\begin{equation}
 \frac{d\sigma}{dt} = \frac{9\pi \alpha_s^2}{\left(t+m_D^2\right) ^2}\left(\frac{1}{2}+\frac{m^2_D}{2 s}\right) 
\label{eq3}
\end{equation}
where $s,t$ are the Mandelstam variables and $m_D=$ 0.7 GeV.

\begin{figure}[ht]
\centering
\includegraphics[height=1.9in,width=2.4in]{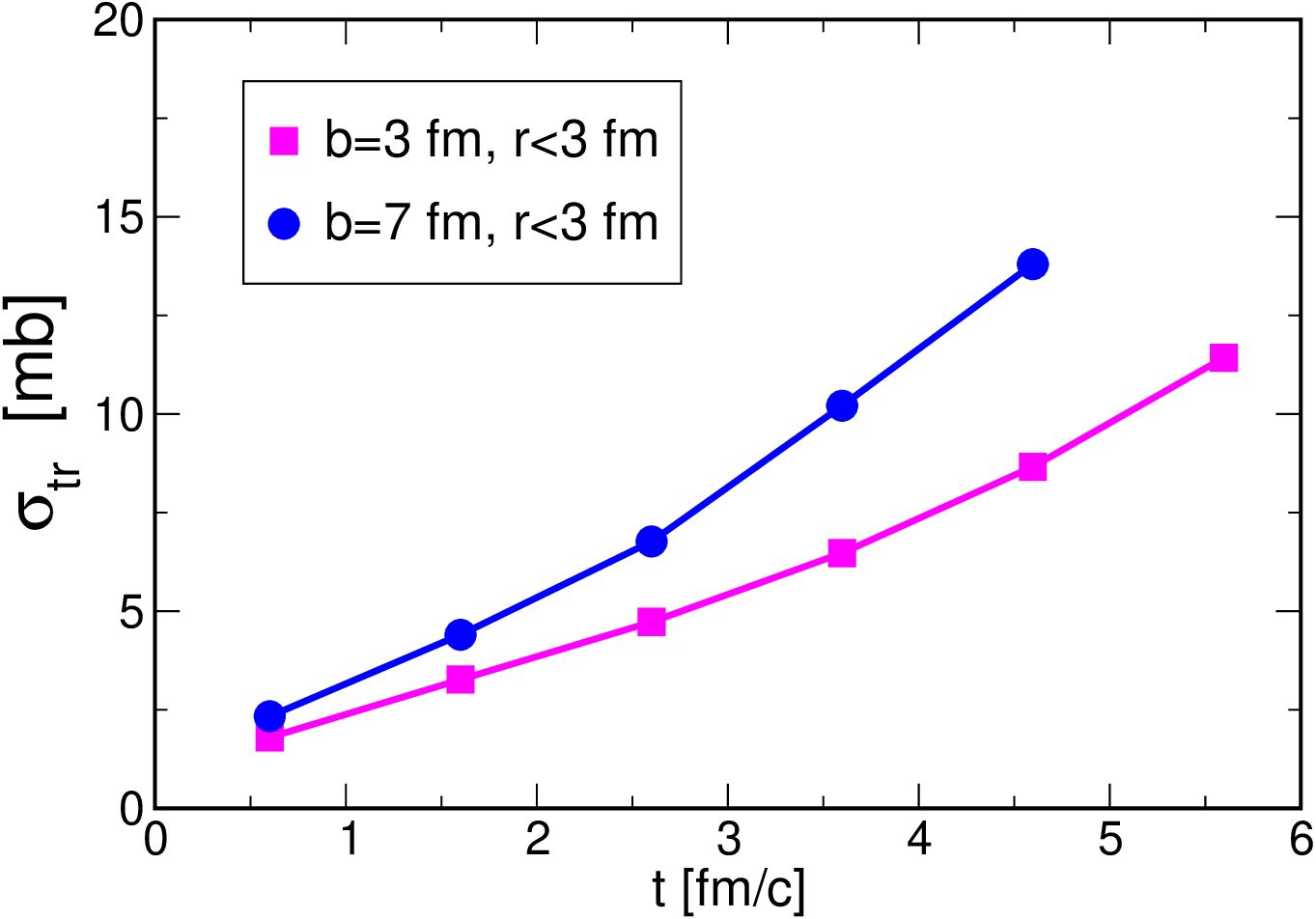}
\hspace{1.3cm}
\includegraphics[height=1.9in,width=2.4in]{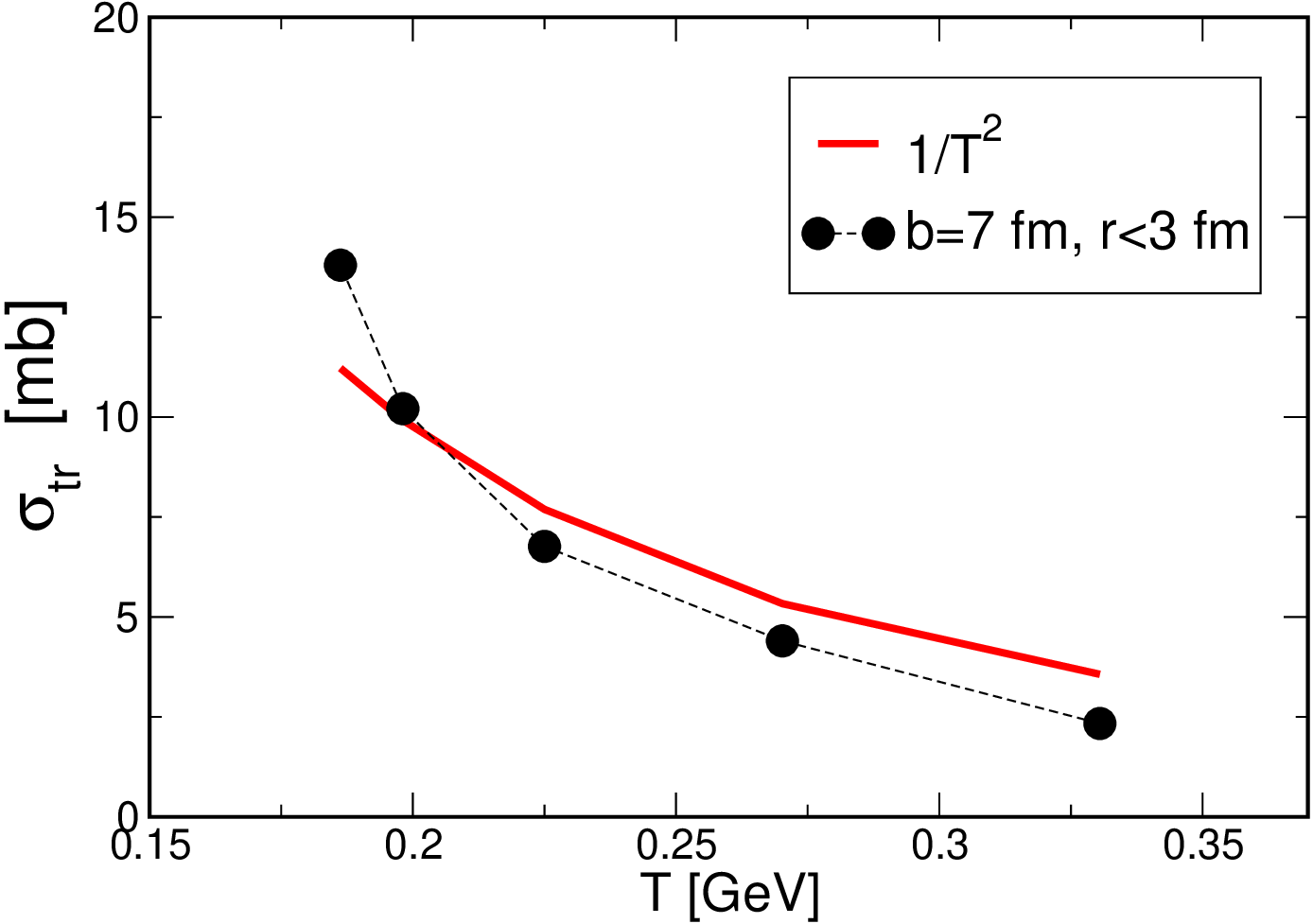}
\caption{Left: Time dependence of the cross section in a central region of rapidity ($|y|<0.2$) and transverse 
radius $r<3$ fm. Right: Temperature dependence of the cross section in the central region compared with
the $T^{-2}$ dependence. }
\label{sigma1}
\end{figure}

Our approach is to artificially keep the $\eta/s$ of the medium constant during the dynamics of the collisions
in a way similar to \cite{Abreu:2007kv,Molnar:2008jw},
but evaluating
locally in space and time the strength of the cross section $\sigma_{tr}(\rho(\textbf{r}),T)$
needed to keep the $\eta/s$ constant. From Eq. (\ref{eq2}) we see that assuming locally the thermal equilibrium
this can be obtained evaluating in each $\alpha$ cell the cross section according to:
\begin{equation}
 \sigma_{tr,\, \alpha} =\frac{4}{15} \frac{\langle p \rangle_{\alpha}}{\rho_{\alpha} (4 - \mu_{\alpha}/T )}\frac{1}{\eta/s}
\label{eq5}
\end{equation}
with $4\pi\eta/s$ set in the range $1-4$. 
We notice that a guideline on the temperature and time dependence of the cross section can be obtained
considering the simple case of a free massless gas for which $s=g \frac{2 \pi^2}{45} T^3$, and therefore
neglecting $\mu$ in Eq. (\ref{eq5}) one gets $\sigma_{tr} \sim T^{-2}$ for $4\pi\eta/s = 1$. Furthermore
a simple Bjorken expansion which means $T\sim \tau^{-1/3}$ gives $\sigma_{tr}\propto \tau^{2/3}$ which is
the approximate prescription adopted in \cite{Abreu:2007kv,Molnar:2008jw}.
In Fig.\ref{sigma1} it is shown $\sigma_{tr}(\tau)$ evaluated locally in space in a cylinder of radius 3 fm as a function
of time, we see on the left the approximate $\tau^{2/3}$ and on the right the agreement with the estimated
$T^{-2}$ behavior. 

\section{Results}
Partons are initially distributed according to the Glauber model, hence also the eccentricity is similar
to the one used in standard calculations. We also start our simulation like in hydrodynamics
at a time $t=0.6$ fm, assuming free-streaming evolution from $t=t_0$ to $t=0.6 fm$. Partons with $p_T < p_0=$2 GeV are distributed according to a thermalized spectrum, while for $p_T >p_0$ we take the spectrum
of non-quenched minijets as calculated in \cite{Zhang:2001ce}.

A first important issue has been to investigate if the scaling behavior of the elliptic flow with
the initial eccentricity $\epsilon_x$ and the system size typical of ideal hydrodynamics \cite{Bhalerao:2005mm}
persists at finite $\eta/s$ close to the lower bound. The interest for such a study is triggered by the
recent observation by PHENIX \cite{Adare:2006ti} of a scaling of $v_2(p_T)/\langle v_2 \rangle$ up to $p_T \sim 3$ GeV, a region usually considered
out of the range where hydrodynamics should work. 
In Ref.\cite{Ferini:2008he} it has been shown that both ${v_2(p_T)}/{\epsilon_x}$ and ${v_2(p_T)}/{\langle v_2\rangle}$
scale with the impact parameter and system size if the fireball is let evolving down to vanishing energy density and a
constant $\eta/s$. This was found for both $Au+Au$ and $Cu+Cu$ at $\sqrt{s}=$ 200 GeV.
This indicates that the scaling $v_2(p_T)/\langle v_2\rangle$, which
is advocated as a signature of the hydrodynamical behavior \cite{Adare:2006ti}, is a more general property that holds also
at finite mean free path or shear viscosity at least for values close to the lower bound. 
Moreover the scaling is shown to persist also at higher $p_T$ ($\sim$ 3 GeV) where
not only the scaling but also the saturation shape is correctly reproduced
by the parton cascade approach. 
We mention that our results are in qualitative agreement with predictions from viscous hydrodynamics \cite{Romatschke:2007mq,Song:2007fn}, 
however we are not aware of an explicit investigation of $v_2(p_T)/\epsilon_x$ and $v_2(p_T)/\langle v_2 \rangle$ scaling within hydrodynamics.

However this first result is in contrast with the experimental observation that ${v_2(p_T)}/{\langle v_2\rangle}$ 
scaling is observed \cite{Adare:2006ti} while the ${v_2(p_T)}/{\epsilon_x}$ is not \cite{:2008ed}. 
However it should be noticed that when the local energy density $\epsilon \sim 1$ GeV/fm$^3$ the QGP phase 
stops and is likely that the $v_2$ does not develops significantly during the hadronic phase as suggested 
also by recent analysis
of experimental data \cite{Afanasiev:2007tv,Abelev:2007rw}. 
In Ref.\cite{Ferini:2008he} for investigating the impact of a QGP freeze-out on the eccentricity scaling we used a cut-off freeze-out stopping the collisions among partons as the local energy density drops below $\epsilon_{f.o.}=$0.7 GeV/fm$^3$ an intermediate value in the range corresponding to a mixed quark-hadron phase \cite{Kolb:2003dz}.

\subsection{Effect of QGP freeze-out on $v_2$ and $v_4$} . 

\begin{figure}[ht]
\vskip0.3cm
\centering
\includegraphics[height=2.0in,width=2.4in]{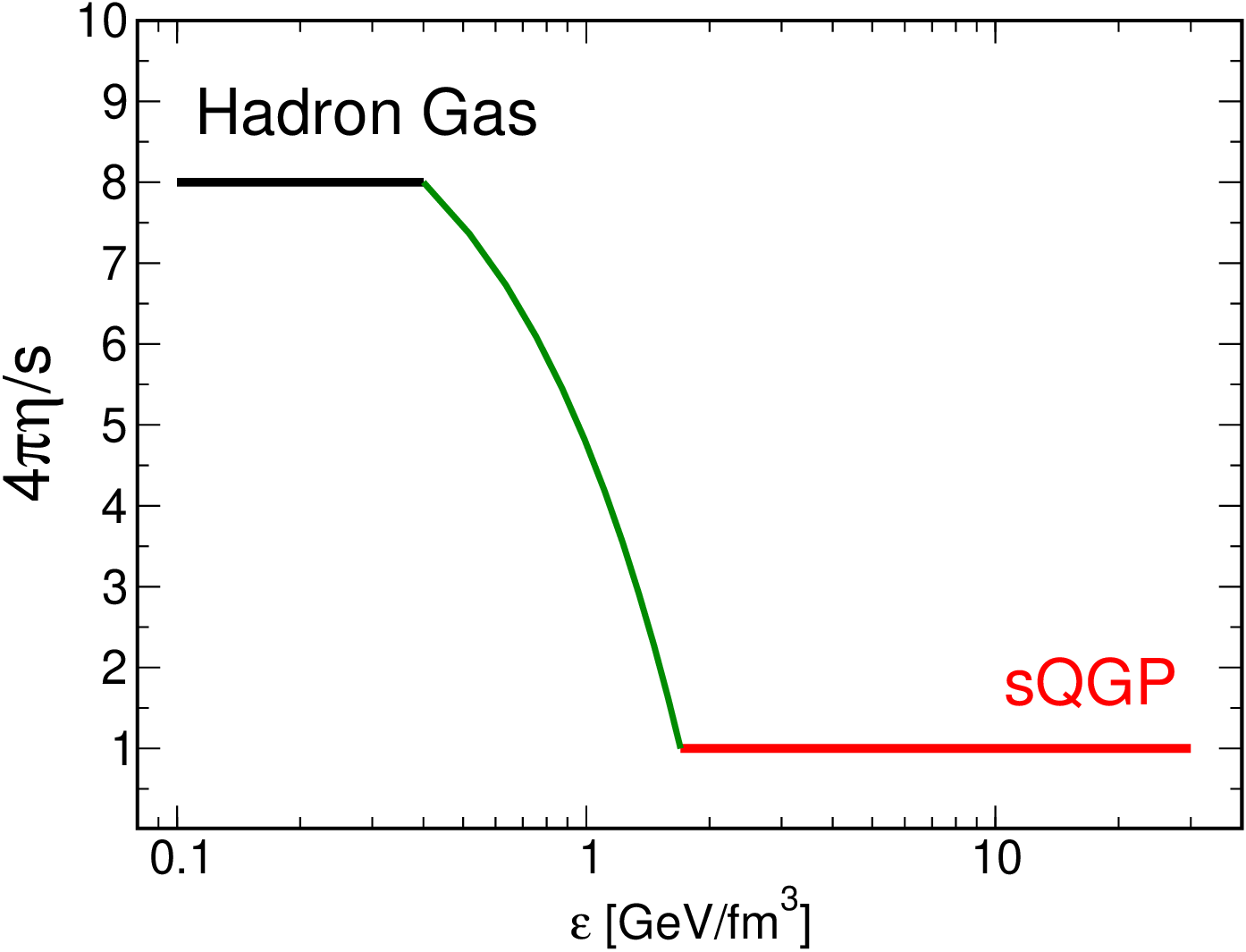}
\hskip0.8cm
\includegraphics[height=2.0in,width=3.8in]{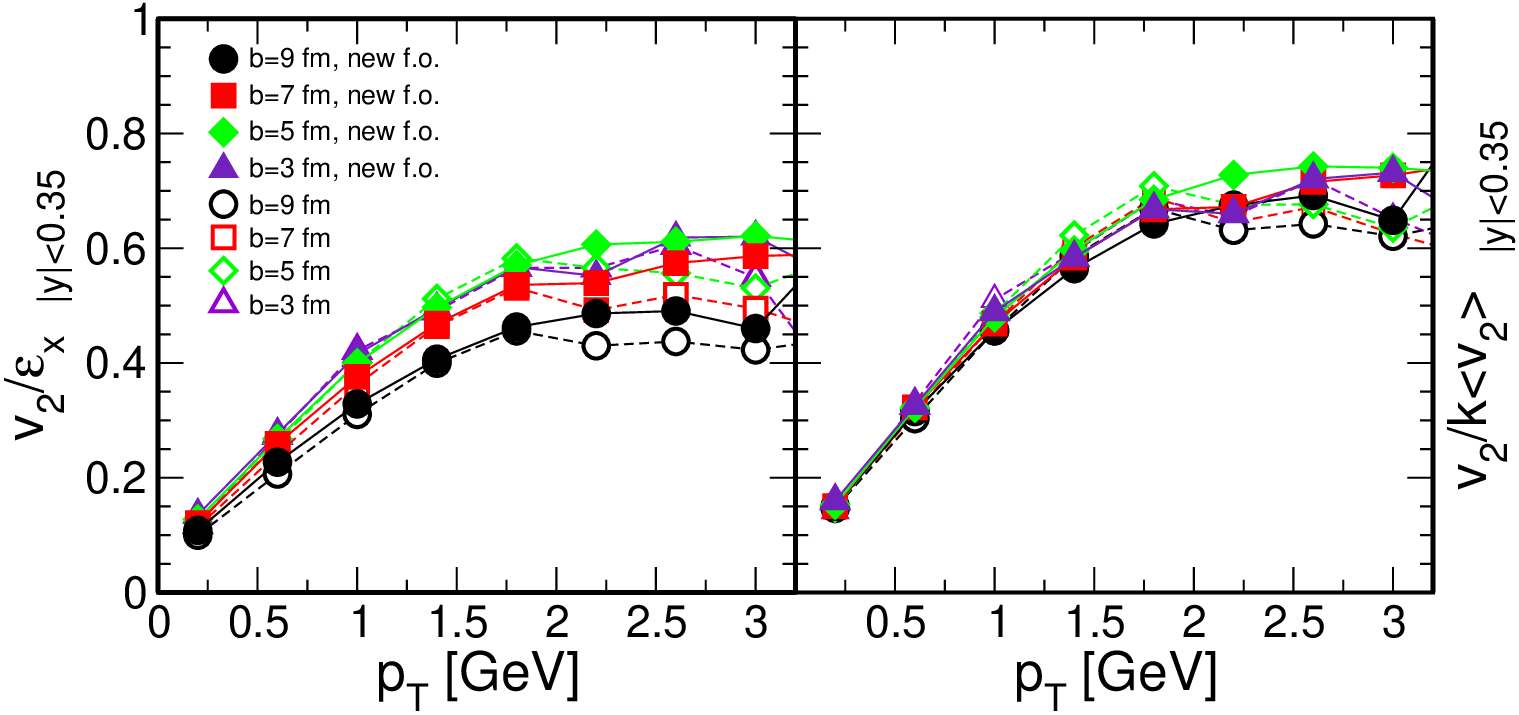}
\caption{Left: Behavior of the $4\pi\eta/s$ versus the energy density in the freeze-out procedures. The value for the hadrons gas is taken from \cite{h-viscosity}. Right: parton $\frac{v_2}{\epsilon_x}$ (left panel)  and $\frac{v_2}{k\langle v_2\rangle}$ (right panel) in the central rapidity region ($|y|<0.35$) for Au+Au at $\sqrt{s}=200 AGeV$. Different symbols refer to cascade simulations at various impact parameters for $4\pi\eta/s=1$. The results
with the cut-off f.o. (dashed lines, open symbols)are compared with the one with the new f.o shown by solid lines
and filled symbols.}
\label{fig11}
\end{figure}

When the freeze-out condition is implemented a sizeable reduction for the 
elliptic flow is observed (see Fig. 2 in Ref.\cite{Ferini:2008he}), especially for the most 
peripheral collisions and at intermediate $p_T$. 
Correspondingly the scaling of elliptic flow with the initial spatial eccentricity is broken 
\cite{Ferini:2008he}. In particular $v_2(p_T)/\epsilon_x$ varies of nearly $40\%$ from b=3 fm to b=9 fm in the 
intermediate $p_T$ region ($\sim$3 GeV). The amount of such a spreading is 
consistent with the data reported by \cite{:2008ed} for the centrality 
selections $0-10\%$ and $10-40\%$, with central collisions exhibiting a 
bigger elliptic flow to eccentricity ratio than the peripheral ones, see also Fig.\ref{fig11}. 
On the other hand, the scaling of $v_2/\langle v_2\rangle$ with the impact parameter is 
still present. We are therefore driven to the conclusion that the breaking of the 
${v_2(p_T)}/{\epsilon}$ scaling traces back to the freeze out physics, which deserves a deeper investigation.

This is partially due to the fact that
${\langle v_2\rangle}/{\epsilon_x}$ does not exactly scale with centrality, as pointed out by
STAR and PHOBOS \cite{Abelev:2007rw,:2008ed} measurements
and also qualitatively confirmed by our approach \cite{Ferini:2008he}.
In hydrodynamics it has also recently been pointed out that the scaling with eccentricity for the ${\langle v_2\rangle}$
is indeed slightly broken by the freeze-out\cite{Song:2008si}. This is roughly speaking in line with our results. 
However there the effect is smaller because in hydrodynamics the viscosity is kept fixed at the minimum
value also in the hadronic phase.

Our previous work drew the attention to the freeze-out (f.o.) but certainly its implementation as a sudden
cut-off f.o. at the fixed $\epsilon_{f.o.}=$ 0.7 GeV/fm$^3$ is an oversimplified approach to the problem. Here we present 
a first improvement that is realized changing the local $\eta/s$ from the conjectured value in the QGP phase to the
one estimated for an hadronic phase \cite{h-viscosity} that is about eight times larger.
The local $\eta/s$ starts to change when $\epsilon <$ 1.7 GeV/fm$^3$ and rises linearly with $\epsilon$ reaching the
values of $4\pi\eta/s=8$ when $\epsilon =$ 0.4 GeV/fm$^3$ as shown in the left panel of
Fig.\ref{fig11} (left). Such a range is the one corresponding to the mixed phase
of the standard equation of state (EOS) used in hydrodynamical calculations \cite{Kolb:2003dz}.
This procedure of course is more realistic respect to the sudden cut-off, however in Fig.\ref{fig11} (right) we can see 
that the results are quite similar in the two f.o. schemes, i.e. with the cut-off at $\epsilon_{f.o.}$ 
(open symbols, dashed line)
and for the new f.o. with the smooth increase of viscosity (filled symbols, solid line). The results of Ref.\cite{Ferini:2008he}
and the relative discussion are therefore confirmed. Of course it is important to understand the physical motivation for the weak
dependence of the $v_2(p_T)$ on the f.o. procedure. Looking at the time scale one can realize that indeed most of the
elliptic flow is formed earlier than the time scale at which the system is at $\epsilon \approx \epsilon_{f.o.}$ and hence
the details of the interaction during such a phase are less relevant.
In this perspective we think that this is also the origin of the moderate impact of different EOS on
$v_2(p_T)$ seen within hydrodynamics in \cite{Bluhm:2007nu}. 

\begin{figure}[ht]
\centering
\includegraphics[height=2.0in,width=2.6in]{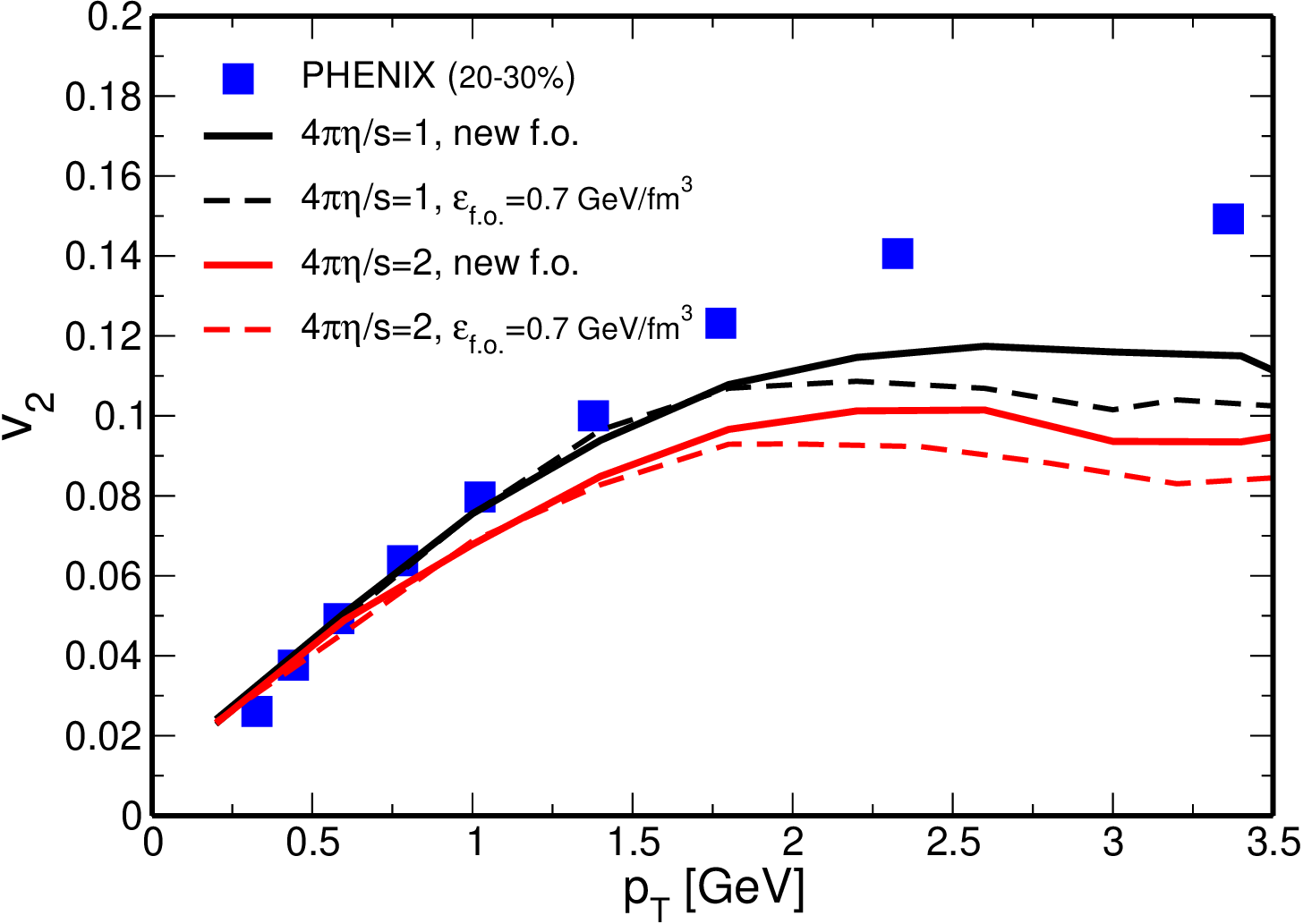}
\hskip1.4cm
\includegraphics[height=2.0in,width=2.6in]{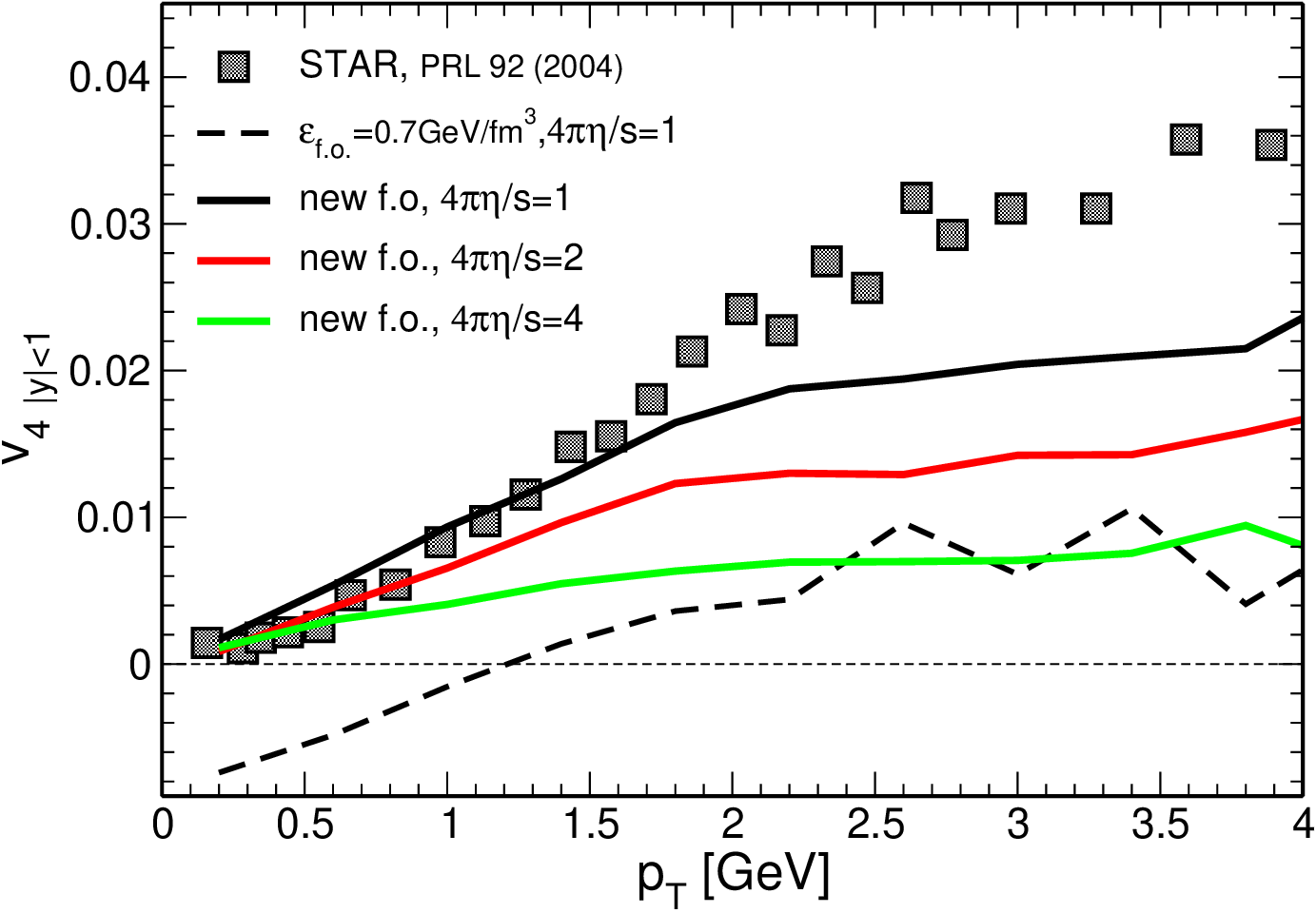}
\caption{Left: Differential elliptic flow for Au+Au collisions $\sqrt{s_{NN}}=$200 AGeV for two values of $\eta/s$ and the two
f.o. schemes: the cut-off at $\epsilon_{f.o.}$ (dashed lines) and the new f.o. (solid line); squares are data from \cite{Adare:2006ti}. Right: $v_4(p_T)$ for three values of $\eta/s$ in the new f.o. procedure (solid line) and 
the cut-off one (dashed line); data are from \cite{Adams:2003zg}.}
\label{fig4}
\end{figure}

The study of azimuthal anisotropy can be extended to higher harmonics and in particular a finite fourth harmonic $v_4 = <cos(\phi)>$ \cite{Adams:2003zg} has been measured at RHIC for the first time in heavy-ion collisions. A first study with AMPT model has shown that it is more sensible to the partonic dynamics \cite{Chen:2004dv}.
We point out that it has a formation time that is delayed respect to the $v_2$ one and therefore it becomes 
much more dependent on the dynamics at energy densities in the region of the mixed phase.
This is clearly shown in Fig.\ref{fig4} where we can see that $v_2(p_T)$ is essentially not affected
by the f.o. scheme while $v_4(p_T)$ drastically depends on it (compare the solid and dashed line both
at $4\pi\eta/s=$1). Moreover only the smooth f.o. procedure is able to reproduce a
$v_4(p_T)$ similar to the one observed in experiments (squares). Therefore we warn about the use of a cut-off f.o. 
as done also by other groups \cite{Xu:2007jv} even if the evaluation of the $v_2(p_T)$ can still be considered quite 
reliable. The observation that $v_4(p_T)$ is formed at energy density $\epsilon \sim$ 0.5-2 GeV/fm$^3$ let us envisage that $v_4$ and not $v_2$ is a good
probe of the EOS in the cross-over region and it would be interesting to examine it also within hydrodynamics.

\begin{figure}[ht]
\centering
\includegraphics[height=2.0in,width=4.5in]{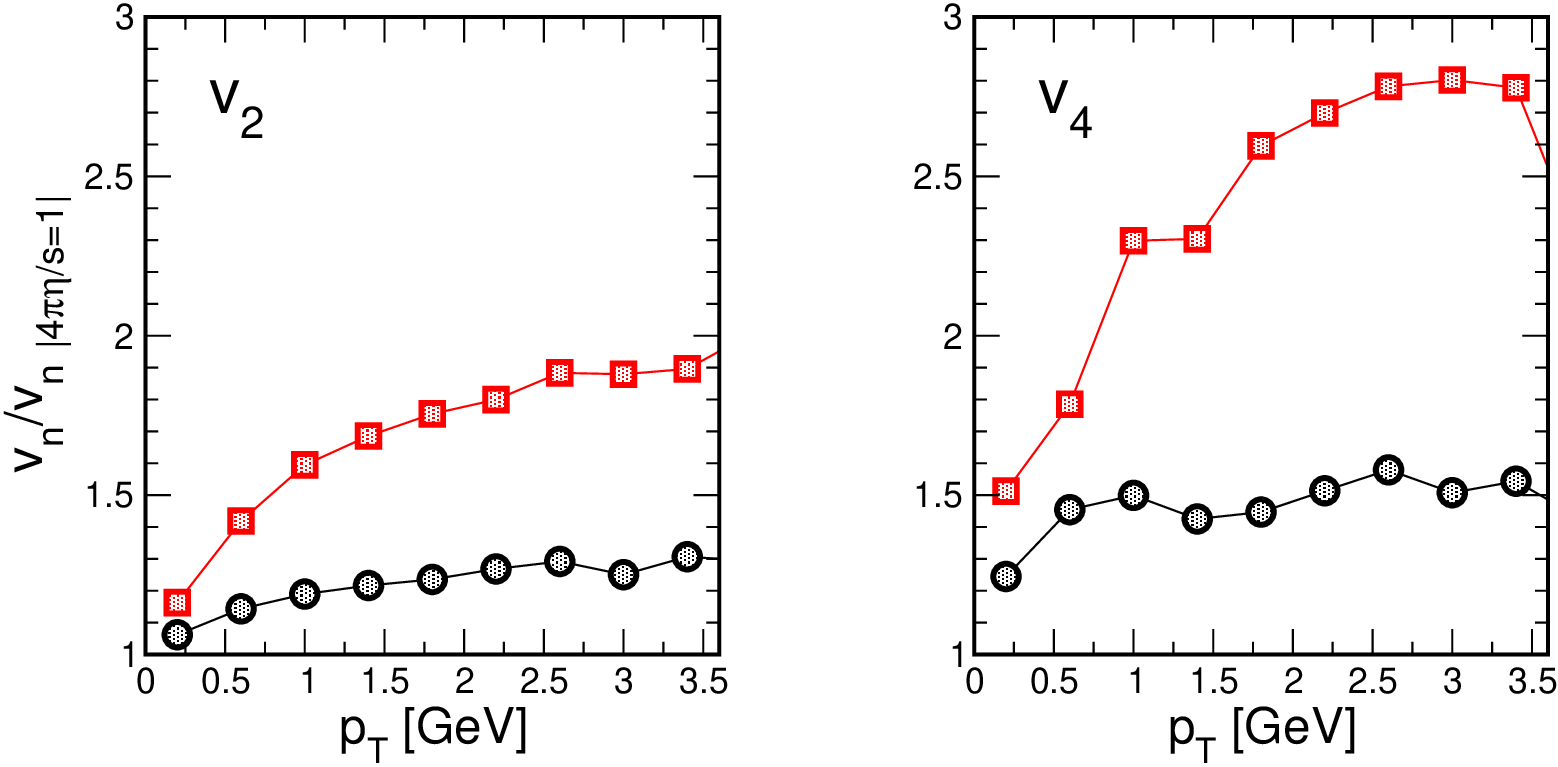}
\caption{Au+Au $\sqrt{s_{NN}}$=200 AGeV and b=7fm. Left: relative enhancement of $v_2(p_T)$ at $4\pi\eta/s=$2 (circles) and 4 (squares) respect to the one at 1. Right the same left panel but for the $v_4$.}
\label{fig5}
\end{figure}

A last point we emphasize is the dependence of $v_4(p_T)$ on the value of $\eta/s$. In Fig.\ref{fig5}
there is a comparison between the relative change in $v_2(p_T)$ (left) and $v_4(p_T)$ (right) when the
$4\pi\eta/s$ is increased by a factor of two and four respect to the lower bound. 
It is seen that the $v_4(p_T)$ is more sensitive 
and therefore can in principle give a better constraint on the $\eta/s$ of the QGP. Moreover it
seems, see Fig.\ref{fig4}, that also the $v_4(p_T)$ confirms that $\eta/s$ is at most a couple of time the lower bound.

Furthermore as for the elliptic flow there is a good agreement with data at low $p_T$ while at intermediate $p_T$ a room
is open for the presence of a coalescence mechanism that would make the estimate of $\eta/s$ consistent with
the observation of quark number scaling for both $v_2$ and $v_4$ \cite{Annual,Greco:2007nu}.

\section{Conclusion}
 
We have investigated the dependence on the shear viscosity and the f.o. dynamics of the elliptic flow $v_2(p_T)$ 
and of the $v_4(p_T)$. 
As a first result we find that the approximate scaling of $v_2(p_T)/\epsilon_x$
advocated as a signature of the perfect hydrodynamical behavior \cite{Adare:2006ti} can still hold also at
finite viscosity up to intermediate $p_T$ \cite{Ferini:2008he}. However such a scaling versus centrality
is present only if one makes the fireball evolve without any f.o. condition. 
We found in \cite{Ferini:2008he} that the freeze-out condition for the partonic dynamics is 
essential to reproduce the sizeable breaking of the $v_2(p_T)/\epsilon_x$ scaling together with a persisting
$v_2(p_T)/\langle v_2\rangle$ scaling. This indicates the relevance of the freeze-out of QGP dynamics.
In a cascade approach it is frequently used a f.o. condition based on the switch-off of the collisions
when the energy density $\epsilon$ is below a fixed value $\epsilon_{f.o.}$. Here we have also studied the effect of a smooth f.o. procedure that takes into account the change of $\eta/s$ going from the QGP to the hadronic matter.
We find that both the f.o. modelling give quite similar  $v_2(p_T)/\epsilon_x$ and $v_2(p_T)/\langle v_2\rangle$,
hence also the estimate of $\eta/s$ based on the $v_2$ obtained with the ``cut-off'' at $\epsilon_{f.o.}$
can be considered quite stable.
On the other hand we found a quite strong dependence of the $v_4(p_T)$ on both the f.o. procedure and the $\eta/s$.
This is due to its formation scale that is close to the life-time of the QGP phase. We find that the 
$\epsilon_{f.o.}$ cut-off is not able at all to reproduce the shape and even the sign of the observed $v_4(p_T)$ while the
more realistic smooth f.o. produces a shape similar to the experimental data. A first estimate shows that also $v_4(p_T)$
indicates a plasma with $4\pi\eta/s \sim 1-2$ as the analysis of $v_2(p_T)$ does \cite{Drescher:2007cd,Romatschke:2007mq,Song:2007fn,Xu:2007jv,Ferini:2008he} which is an encouraging consistency.

As a last point we point out that at intermediate $p_T$ there are several evidences
for hadronization via coalescence and it has been shown that due to a coalescence mechanism the parton $v_2$ translates into a nearly doubled hadron $v_2$ \cite{Annual}. On the other hand the f.o. condition in a parton cascade approach
generates a shape of $v_{2,4}(p_T)$ that agrees with the data at low $p_T$ and leaves the room for a coalescence
enhancement at intermediate $p_T$.
Therefore a definite evaluation of $\eta/s$ from $v_2(p_T)$ and/or $v_4(p_T)$ needs 
a further development of the parton cascade approach that includes self-consistently
the coalescence and fragmentation process
in order to account also for the baryon-meson quark number scaling. 
Finally a quantitative estimate should also include the
effect of a non vanishing $\epsilon-3p$ in the cross-over transition and is known to lower the sound velocity 
and hence the $v_2$ \cite{Song:2007fn,Bhalerao:2005mm}.


\begin{thebibliography}{99}
\itemsep -2pt 
\bibitem{Kolb:2003dz}
  P.~F.~Kolb and U.~W.~Heinz, arXiv:nucl-th/0305084;  P. Houvinen,
   arXiv:nucl-th/0305064;
    in {\it Quark Gluon Plasma 3}, R.C.
    Hwa and X.N. Wang (Eds.), World Scientific, Singapore, 2004.

\bibitem{Ackermann:2000tr}
  K.~H.~Ackermann {\it et al.}  [STAR Collaboration],
  Phys.\ Rev.\ Lett.\  {\bf 86} (2001) 402

\bibitem{Abelev:2007rw}
  B.~I.~Abelev {\it et al.}  [STAR Collaboration],
  Phys.\ Rev.\ Lett.\  {\bf 99} (2007) 112301

\bibitem{greco}V. Greco, C.M. Ko, and P. L\'evai, Phys. Rev. Lett. \textbf{90}, 202302 (2003); Phys. Rev. C \textbf{68}, 034904 (2003); R.C. Hwa and C.B. Yang, Phys. Rev. C {\bf 67}, 034902 (2003); R.J. Fries, B. M\"uller, C. Nonaka, and S.A. Bass, Phys.Rev. Lett. \textbf{90}, 202303 (2003); 
Phys. Rev. C {\bf 68}, 044902 (2003); D.~Molnar and S.A.~Voloshin, Phys. Rev. Lett. {\bf 91}, 092301 (2003).

\bibitem{Annual} R.J. Fries, V. Greco, P. Sorensen, 
  Ann. Rev. Nucl. Part. Sci., \textbf{58} (2008) 177;
  arXiv:0807.4939 [nucl-th]

\bibitem{Kovtun:2004de}
  P.~Kovtun, D.~T.~Son and A.~O.~Starinets,
  Phys.\ Rev.\ Lett.\  {\bf 94} (2005) 111601

\bibitem{Drescher:2007cd}
  H.~J.~Drescher, A.~Dumitru, C.~Gombeaud and J.~Y.~Ollitrault,
  Phys.\ Rev.\  C {\bf 76} (2007) 024905


\bibitem{Romatschke:2007mq}
  P.~Romatschke and U.~Romatschke,
  Phys.\ Rev.\ Lett.\  {\bf 99} (2007) 172301

\bibitem{Song:2007fn}
  H.~Song and U.~W.~Heinz,
  Phys.\ Lett.\  B {\bf 658} (2008) 279; 

\bibitem{Song:2008si}
  H.~Song and U.~W.~Heinz,
  Phys.\ Rev.\  C {\bf 78} (2008) 024902


\bibitem{Xu:2007jv}
  Z.~Xu, C.~Greiner and H.~Stocker,
  Phys.\ Rev.\ Lett.\  {\bf 101} (2008) 082302



\bibitem{Ferini:2008he}
  G.~Ferini, M.~Colonna, M.~Di Toro and V.~Greco,
  Phys. Lett. \textbf{B} (2008)
  arXiv:0805.4814 [nucl-th].

\bibitem{Molnar:2008jw}
  D.~Molnar,
  arXiv:0806.0026 [nucl-th].

\bibitem{Abreu:2007kv}
  N.~Armesto {\it et al.},
  J.\ Phys.\ G {\bf 35} (2008) 054001

\bibitem{Bhalerao:2005mm}
  R.~S.~Bhalerao, J.~P.~Blaizot, N.~Borghini and J.~Y.~Ollitrault,
  Phys.\ Lett.\  B {\bf 627} (2005) 49

\bibitem{Adare:2006ti}
  A.~Adare {\it et al.}  [PHENIX Collaboration],
  Phys.\ Rev.\ Lett.\  {\bf 98} (2007) 162301

\bibitem{:2008ed}
  B.~I.~Abelev {\it et al.}  [STAR Collaboration],
  Phys.\ Rev.\  C {\bf 77} (2008) 054901


\bibitem{Xu:2004mz}
  Z.~Xu and C.~Greiner,
  Phys.\ Rev.\  C {\bf 71} (2005) 064901

\bibitem{degroot} S.R. De Groot et al., \textit{Relativistic Kinetic Theory}, North-Holland, Amsterdam,
1980.

\bibitem{moln02}D. Molnar, and M. Gyulassy, Nucl. Phys.{\bf A697} (2002) 495; 
Erratum in Nucl. Phys. {\bf A703} (2002) 893.

\bibitem{Zhang:2001ce}
  Y.~Zhang, G.~I.~Fai, G.~Papp, G.~G.~Barnafoldi and P.~Levai,
  Phys.\ Rev.\  C {\bf 65} (2002) 034903

\bibitem{h-viscosity}
 M. Prakash, R. Venugopalan, and G. Welke, Phys. Rept. 227, 321 (1993);
J.-W. Chen and E. Nakano, Phys. Lett. B647, 371 (2007).

\bibitem{Afanasiev:2007tv}
  S.~Afanasiev {\it et al.}  [PHENIX Collaboration],
  Phys.\ Rev.\ Lett.\  {\bf 99} (2007) 052301

\bibitem{Bluhm:2007nu}
  M.~Bluhm, B.~Kampfer, R.~Schulze, D.~Seipt and U.~Heinz,
  Phys.\ Rev.\  C {\bf 76} (2007) 034901

\bibitem{Greco:2007nu}
  V.~Greco,
  Eur.\ Phys.\ J.\ ST {\bf 155} (2008) 45

\bibitem{Adams:2003zg}
  J.~Adams {\it et al.}  [STAR Collaboration],
  Phys.\ Rev.\ Lett.\  {\bf 92} (2004) 062301

\bibitem{Chen:2004dv}
  L.~W.~Chen, C.~M.~Ko and Z.~W.~Lin,
  Phys.\ Rev.\  C {\bf 69} (2004) 031901


\bibitem{Kolb:2004gi}
  P.~F.~Kolb, L.~W.~Chen, V.~Greco and C.~M.~Ko,
  Phys.\ Rev.\  C {\bf 69} (2004) 051901


\end{thebibliography}
\end{document}